\documentclass[sigconf]{acmart}

\usepackage{booktabs}

\usepackage{tablefootnote}
\usepackage{mathtools}
\usepackage{balance}
\usepackage{amssymb,amsmath}
\usepackage{proof,latexsym}
\usepackage{listings}
\usepackage{tcolorbox}
\usepackage{enumerate}
\usepackage{ifthen}
\usepackage{amssymb}
\usepackage{multicol}
\usepackage{multirow}
\usepackage[ruled]{algorithm2e}
\usepackage{enumitem}
\usepackage{todonotes}

\usepackage[para]{threeparttablex}

\setcopyright{none}
\renewcommand\footnotetextcopyrightpermission[1]{}

\begin{document}
\title{On Reliability of Patch Correctness Assessment}


\author{Xuan-Bach D. Le}
\affiliation{%
  \institution{CyLab Security and Privacy Institute, Carnegie Mellon University}
  \city{Moffett Field, CA, USA}
}
\email{bachldx@cmu.edu}

\author{Lingfeng Bao}
\affiliation{%
  \institution{College of Computer Science and Technology, Zhejiang University}
  \state{China}
}
\email{lingfengbao@zju.edu.cn}

\author{David Lo}
\affiliation{%
  \institution{School of Information Systems, Singapore Management University}
  \city{Singapore}
  }
\email{davidlo@smu.edu.sg}

\author{Xin Xia}
\affiliation{%
  \institution{Faculty of Information Technology, Monash University}
  \state{Australia}
}
\email{xin.xia@monash.edu}

\author{Shanping Li}
\affiliation{%
  \institution{College of Computer Science and Technology, Zhejiang University}
  \state{China}
}
\email{shan@zju.edu.cn}

\begin{abstract}
Current state-of-the-art automatic software repair (ASR) techniques rely heavily on incomplete specifications, e.g., test suites, to generate repairs. This, however, may render ASR tools to generate incorrect repairs that do not generalize. To assess patch correctness, researchers have been following two typical ways separately: (1) Automated annotation, wherein patches are
automatically labeled by an independent test suite (ITS) -- a patch passing the ITS is regarded as correct or generalizable, and incorrect otherwise, (2) Author annotation, wherein authors of ASR techniques annotate correctness labels of patches
generated by their and competing tools by themselves. While automated annotation fails to prove that a patch is actually correct,
author annotation is prone to subjectivity. This concern has caused an on-going debate on appropriate ways to assess the effectiveness of numerous ASR techniques proposed recently.

To address this concern, we propose to assess reliability of author and automated annotations on patch correctness assessment.
We do this by first constructing a gold set of correctness labels for 189 randomly selected patches generated by 8 state-of-the-art
ASR techniques through a user study involving 35 professional developers as independent annotators. By measuring inter-rater
agreement as a proxy for annotation quality -- as commonly done in the literature -- we demonstrate that our constructed gold set is on par with other high-quality gold sets. We then compare labels generated by author and automated annotations with this gold set to assess reliability of the patch assessment methodologies. We subsequently report several findings and highlight implications for future studies.
\end{abstract}

\maketitle

\section{Introduction}


Bug fixing is notoriously difficult, time-consuming, and costly~\cite{NIST,Britton13}. Hence, effective automatic software repair (ASR) techniques that can help reduce the onerous burden of this task, is of tremendous value. Interest in ASR has intensified as demonstrated by substantial recent work devoted to the
area~\cite{mechtaev2015directfix,mechtaev2016angelix,acs:xiong2016precise,long2015staged,long2016automatic,nopol:xuan,le2012systematic,kim2013automatic,le2015should, deductiverepair, lejfix2017, chandra2011angelic}, bringing the futuristic idea of ASR closer to reality. ASR can be generally divided into two main families including heuristics- vs semantics-based approaches, classified by the way they generate and traverse the search space for repairs. 

Traditionally, test cases are used as the primary criteria for correctness judgment of machine-generated patches -- a patch is deemed as \emph{correct} if it passes all tests used for repair~\cite{le2012systematic}. This assessment methodology, however, has been shown to be ineffective as there could be multiple patches passing all tests but are still indeed \emph{incorrect}~\cite{qi2015analysis, long2016analysis}. Although the search space of ASR varies depending on the nature of underlying techniques, it is often huge and contains many plausible repairs, which unduly pass all tests but fail to generalize to the expected behaviours. This problem, which is often referred to as patch overfitting~\cite{smith2015cure, leoverfitting}, motivates the need of new methodologies to assess patch correctness. The new methodologies need to rely on additional criteria instead of using the test suite used for generating repair candidates (aka. {\em repair test suite}) alone.


To address this pressing concern, most recent works have been following two methods for patch correctness assessment separately:

\begin{itemize}[leftmargin=*]
\item {\bf Automated annotation by independent test suite.} Independent test suites obtained via an automatic test case generation tool are used to determine correctness label of a patch -- see for example~\cite{smith2015cure,le2016empirical}. Following this method, a patch is deemed as \emph{correct} or \emph{generalizable} if it passes both the repair and independent test suites, and \emph{incorrect} otherwise.
\item {\bf Author annotation.} Authors of ASR techniques manually check correctness labels of patches generated by their own and competing tools -- see for example~\cite{xiong2017precise,liu2017identifying}. Following this method, a patch is deemed as \emph{correct} if authors perceive semantic equivalence between generated patches and original developer patches.\vspace{0.2cm}
\end{itemize}

\noindent While the former is incomplete, in the sense that it fails to prove that a patch is actually correct, the latter is prone to author bias. In fact, these inherent disadvantages of the methods have caused an on-going debate as to which method is better for assessing the effectiveness of various ASR techniques being proposed recently. Unfortunately, there has been no extensive study that objectively assesses the two patch validation methods and provides insights into how the evaluation of ASR's effectiveness should be conducted in the future.

This study is conducted to address this gap in research. We start by creating a gold set of correctness labels for a collection of ASR generated patches, and subsequently use it to assess reliability of labels created through author and automated annotations. We study a total of 189 patches generated by 8 popular ASR techniques (ACS~\cite{xiong2017precise}, Kali~\cite{qi2015analysis}, GenProg~\cite{xiong2017precise}, Nopol~\cite{nopol:xuan}, S3~\cite{le2017s3}, Angelix~\cite{mechtaev2016angelix}, and Enumerative and CVC4 embedded in JFix~\cite{lejfix2017}). These patches are for buggy versions of 13 real-world projects, of which six projects are from Defects4J~\cite{just2014defects4j} (Math, Lang, Chart, Closure, Mockito, and Time) and seven projects are from S3's dataset~\cite{le2017s3} (JFlex, Fyodor, Natty, Molgenis, RTree, SimpleFlatMapper, GraphHoper). To determine correctness of each patch, we follow best practice by involving multiple independent annotators in a user study. Our user study involves 35 professional developers; each ASR-generated patch is labeled by five developers by comparing the patch with its corresponding ground truth patch created by the original developer(s) who fixed the bug. By analyzing the created gold set and comparing it with labels generated by three groups of ASR tool authors~\cite{DBLP:journals/ese/MartinezDSXM17, liu2017identifying, le2017s3} and  two automatic test case generation tools such as \textsc{DiffTGen}~\cite{xin2017identifying} and \textsc{Randoop}~\cite{DBLP:conf/icse/PachecoLEB07}, we seek to answer three research questions:

\vspace{0.2cm}\begin{itemize}
\item[{\bf RQ1}] {\em Can independent annotators agree on patch correctness?}
\item[{\bf RQ2}] {\em How reliable are patch correctness labels generated by author annotation?}
\item[{\bf RQ3}] {\em How reliable are patch correctness labels inferred through automatically generated independent test suite?}\vspace{0.2cm}
\end{itemize}

\noindent In RQ1, by measuring inter-rater agreement as a proxy of annotation quality -- as commonly done in the literature~\cite{christopher2008introduction,damessie2017gauging} -- we demonstrate that our gold set is on par with other high-quality gold sets. In the subsequent two RQs, we investigate the strengths and deficiencies of author and automated patch correctness annotation.

We summarize our contributions below:

\begin{itemize}[leftmargin=*]

\item We are the first to investigate the reliability of author and automated annotation for assessing patch correctness. To perform such assessment, we have created a gold set of labelled patches created by a user study involving 35 professional developers. By means with this gold set, we highlight strengths and deficiencies of popular assessment methods employed by existing ASR studies.
     
\item Based on implications of our findings, we provide several recommendations for future ASR studies to better deal with patch correctness validation. Especially, we find that automated annotation, despite being less effective as compared to author annotation, can be used to augment author annotation and reduce the cost of manual patch correctness assessment. \vspace{0.2cm}    
\end{itemize}


The rest of the paper is organized as follows. Section~\ref{sec:background} presents more information on various ASR techniques, existing methods used for patch correctness assessment, and best practice in gold set creation. Next, we describe details of our user study to collect gold set of patch correctness labels in Section~\ref{sec:setup}. Subsequently, we answer RQ1, RQ2, and RQ3 to assess the quality of our gold set, author annotation, and automated annotation in Section~\ref{sec:rq1},~\ref{sec:assessauthor}, and~\ref{sec:assessautomated} respectively. Section~\ref{sec:discussion} discusses implications of our findings, our post-study survey, and threats to validity. We conclude and briefly describe future work in Section~\ref{sec:conclusion}.

\section{Background}\label{sec:background}

In this section, we first present more information about automated software repair (ASR) techniques used in our experiments, including GenProg~\cite{le2012systematic}, Kali~\cite{qi2015analysis}, Nopol~\cite{nopol:xuan}, ACS~\cite{xiong2017precise}, S3~\cite{le2017s3}, Angelix~\cite{mechtaev2016angelix}, and Enumerative and CVC4 embedded in JFix~\cite{lejfix2017}. We subsequently elaborate methods that have been used for assessing patch correctness in ASR research. Finally, we discuss best practice in building gold sets.


\vspace{0.2cm}\noindent\textbf{ASR techniques:} GenProg~\cite{le2012systematic} is one of the first ASR techniques that sparks interests in ASR. Given a buggy program and a set of test cases, at least one of which is failing, GenProg uses a number of mutation operators, such as statement deletion, insertion, and append to create a large pool of repair candidates. It then uses genetic programming to evolve the buggy program until a candidate passing all tests is found. Kali~\cite{qi2015analysis} is a naive ASR technique, which just blindly deletes any statements that are identified as potentially buggy. Despite being simple, Kali has been shown to be as effective and efficient as GenProg. Nopol~\cite{nopol:xuan} is a recently developed ASR technique that focuses on only repairing defective \emph{if-conditions}. Nopol attempts to synthesize an if-condition expression that renders all tests to pass by using program synthesis. In a similar vein, ACS~\cite{xiong2017precise} also focuses on synthesizing repairs for buggy if-conditions. Like Nopol, ACS also uses program synthesis to synthesize repairs. Unlike Nopol, ACS attempts to rank the fix candidates using various ranking functions. Angelix~\cite{mechtaev2016angelix}, S3~\cite{le2017s3}, and JFix~\cite{lejfix2017} use symbolic execution to infer specifications and various program synthesis techniques to synthesize repairs conforming to the inferred specifications.

\vspace{0.2cm}\noindent\textbf{Evaluation of ASR Generated Patches:} Traditionally, test cases are used as the sole criteria for judging correctness of machine-generated patches. By relying on the assumption that a patch that passes the repair test suite is regarded as correct, early repair techniques such as GenProg~\cite{le2012systematic}, AE~\cite{weimer2013leveraging}, and RSRepair~\cite{qi2014strength} reported to produce many such correct patches. However, it has been shown in recent studies that this assumption does not hold true in practice since such patches that pass the repair test suite are indeed still incorrect~\cite{qi2015analysis, long2016analysis}. This shows that repair test suite alone is a weak proxy for assessing patch correctness.

Motivated by the above concern, recent works have thus employed new methods to assess patch correctness: (1) Author annotation, in which authors of repair techniques manually check the correctness of patches generated by their and competing tools by themselves, see for example~\cite{xiong2017precise, le2017s3}; (2) Automated annotation by independent test suite (ITS) generated by automatic test case generation tool, see for example~\cite{smith2015cure, le2016empirical}. Both methods assume that a reference (correct) implementation of the buggy program, which is used as a basis for comparison, is available. Since most ASR techniques try to fix real buggy versions of real programs, the reference implementations can be found in the version control systems of the corresponding projects.


Early work that uses automated annotation by automatically-generated ITS, e.g.,~\cite{smith2015cure}, uses general-purpose automatic test generation tool such as KLEE~\cite{cadar2008klee} to generate an ITS that maximizes the coverage of the reference implementation written in C programming language. To automatically generate test cases for Java programs, \textsc{Randoop}~\cite{DBLP:conf/icse/PachecoLEB07} can be used to randomly generate sequences of method calls that create and mutate objects,
plus an assertion about the result of a final method
call. Recently, Xin \textit{et al.} proposed \textsc{DiffTGen}, a test generation tool for Java programs specifically designed to generate tests that can identify incorrect patches generated by ASR tools~\cite{xin2017identifying}. \textsc{DiffTGen} attempts to generate test cases that cover the syntactic and semantic differences between the machine-patched and human-patched programs. If there are any such test cases that expose the differences in outputs of the programs, the machine-generated patch is deemed as \emph{incorrect} since it results in a different output as compared to the corresponding ground truth human-patched program. \textsc{DiffTGen} has been shown to be able to identify incorrect patches produced by various state-of-the-art ASR tools such as GenProg~\cite{le2012systematic}, Kali~\cite{qi2015analysis}, Nopol~\cite{nopol:xuan}, and HDRepair~\cite{le2016history}.



\vspace{0.2cm} \noindent {\bf Best practice in building gold set:} To build gold set objectively, a common approach is to employ many independent annotators and measure inter-rater agreement as proxy for annotation quality~\cite{christopher2008introduction,Dybkjaer:2007}. Information retrieval (IR) community, especially through the Text REtrieval Conference (TREC)\footnote{\url{http://trec.nist.gov/}}, has employed many annotators through large scale collaborative effort to annotate many document corpora for various retrieval tasks. Many past software engineering studies have also involved independent annotators to construct gold sets. Based on the nature of various tasks, annotators include non-authors who could be undergraduate/graduate students~\cite{rastkar2010summarizing, gachechiladze2017anger,buse2010learning, de2014labeling, zou2015learning} or professional developers~\cite{ormandjieva2007toward, treude2015extracting, rastkar2010summarizing}.

\section{User Study}\label{sec:setup}

We conducted a user study with 35 professional developers to collect correctness labels of patches. In this study, every developer is required to complete several tasks by judging whether patches generated by ASR tools are semantically equivalent to ground truth human patches.

\vspace{0.2cm}\noindent{\bf Patch Dataset.} Since the eventual goal of our study is to assess reliability of author and automated annotations, we need a set of patches that have been labeled before by ASR tool authors and can be used as input to automated test case generation tools designed for program repair. We find the sets of patches recently released by Liu et al.~\cite{liu2017identifying}, Martinez et al.~\cite{DBLP:journals/ese/MartinezDSXM17}, and Le et al.~\cite{le2017s3} to be suitable. Liu et al. and Martinez et al. label a set of 210 patches generated by ASR tools designed by their research groups (i.e., ACS~\cite{xiong2017precise}, and Nopol~\cite{nopol:xuan}) and their competitors (i.e., GenProg~\cite{le2012systematic}, Kali~\cite{qi2015analysis}). Le et al. label a set of 79 patches generated by their ASR tool (i.e., S3~\cite{le2017s3}) and their competitors (i.e., Angelix~\cite{mechtaev2016angelix}, and Enumerative and CVC4 embedded in JFix~\cite{lejfix2017}). The authors label these patches by manually comparing them with ground truth patches obtained from version control systems of the corresponding buggy subject programs.\footnote{Since authors of~\cite{liu2017identifying} and~\cite{xiong2017precise} overlap, we can use the labels to evaluate reliability of author labelling.} These patches can be used as input to \textsc{DiffTGen}, which is a state-of-the-art test generation tool specifically designed to evaluate patch correctness~\cite{xin2017identifying}, and \textsc{Randoop} -- a popular general purpose test case generation tool~\cite{DBLP:conf/icse/PachecoLEB07}.

\begin{table}\vspace{-0.2cm}
\caption{Selected Patches and their Author Label}\vspace{-0.3cm}
\begin{center}
{\small
  \tabcolsep 2.5pt
    \def\arraystretch{0.80}
\begin{tabular}{l|rrrrrrrr}
\toprule
          &\textbf{GenProg} &
          \textbf{Kali} &  \textbf{Nopol} & \textbf{ACS} & \textbf{S3} & \textbf{Angelix} & \textbf{Enum}& \textbf{CVC4}\\

 \midrule
  Incorrect      & 14 & 14 & 84 & 4 & 0 & 7 & 6 & 6\\
  Correct        & 4 & 1 & 6 & 14 & 10 & 2 & 4 & 4\\
  Unknown        & 2 & 2 & 5 & 0 & 0 & 0 & 0 & 0\\
  Total        & 20 & 17 & 95 & 18 & 10 & 9 & 10 & 10\\
 \bottomrule
 \end{tabular}}
 \end{center}
 \vspace{-0.4cm}\label{tbl:dataset}
 \end{table}

Due to resource constraint, i.e., only 35 professional developers agree to spend an hour of their time in this user study, we cut down the dataset to 189 patches by randomly selecting these patches from their original datasets. Details of the dataset of 189 patches are shown in Table~\ref{tbl:dataset}.

\vspace{0.2cm}\noindent{\bf Task Design.} At the start of the experiment, every participant is required to read a tutorial that briefly explains automated program repair and what they need to do to complete the tasks. Afterwards, they can complete the tasks one-by-one through a web interface.

\begin{figure*}[!htb]\vspace{-0.2cm}
\centering
\includegraphics[width=\textwidth,draft=false]{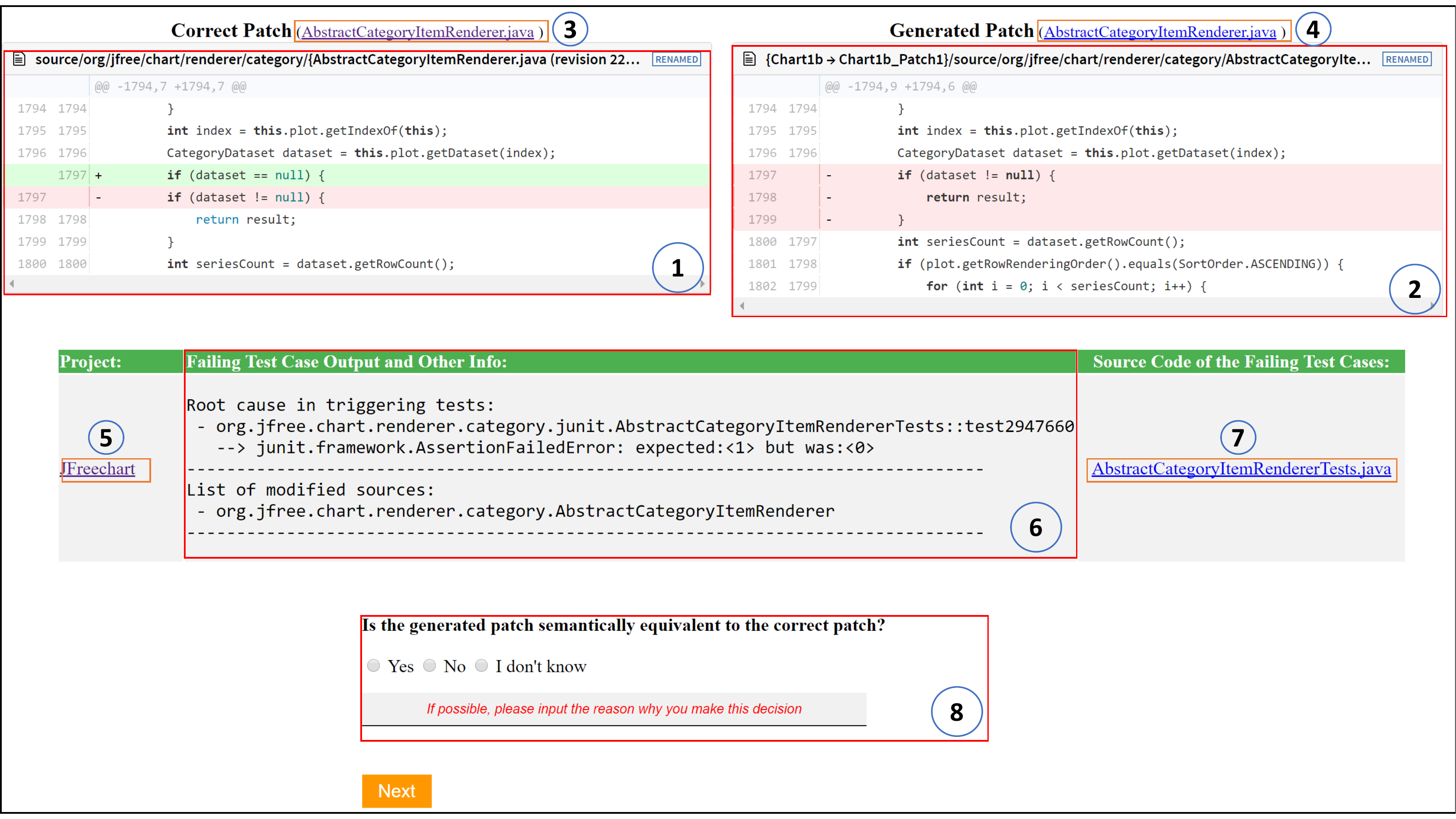}\vspace{-0.2cm}
 \caption{A sample task viewed through our web interface. (1) and (2) are the correct patch and the patch generated by an ASR tool; (3) and (4) are the links to source code files that contain the patches; (5) is the link to the corresponding project's GitHub repository; (6) and (7) are the output of the failed test cases and their source files; (8) is the question we asked a participant to answer.}
 \label{fig:screenshot}
 \vspace{-0.3cm}
\end{figure*}

Figure~\ref{fig:screenshot} shows the screenshot of an example task that we give to our user study participants through a web interface.
For each task, we provide a ground truth patch taken from the version control system of the corresponding buggy subject program, along with a patch that is generated by an automated program repair tool. We also provide additional resources including full source code files that are repaired by the patch, link to the {\sc GitHub} repository of the project, outputs of failing test cases\footnote{These information is generated using Defects4J~\cite{just2014defects4j} info command.}, and source code of the failing test cases. 
Based on this information, participants are asked to evaluate the correctness of the patch by answering the question: \textit{Is the generated patch semantically equivalent to the correct patch?} To answer this question, participants can choose one of the following options: ``Yes'', ``No'' or ``I don't know''. Finally, if they wish to, they can provide some reasons that explain their decision. Our web interface will record participants' answers and the amount of time they need to complete each task.


\vspace{0.2cm}\noindent{\bf Participants and Task Assignment.} Thirty three of the 35 professional developers participating in this study work for two large software development companies (named Company C1 and C2), while another two work as engineers for an educational institution. Company C1 currently has more than 500 employees and Company C2 has more than 2000 employees. Both companies have a large number of active projects that expose developers to various business knowledge and software engineering techniques. All the 35 developers work for projects that use Java as the main programming language.

\begin{figure}
\centering
\includegraphics[width=0.25\textwidth,draft=false]{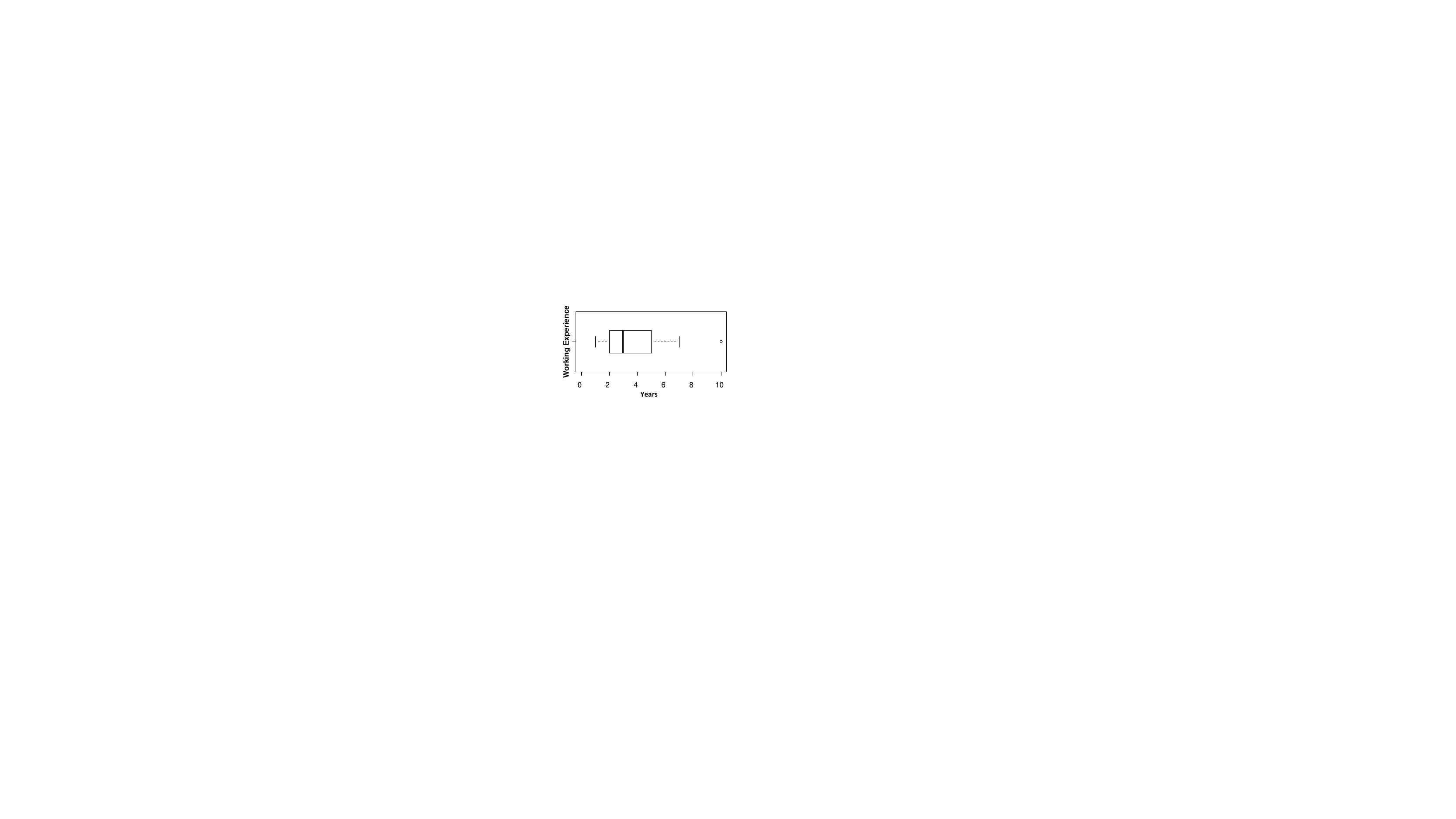}
 \caption{Distribution of participant work experience}
 \label{fig:developer}
 \vspace{-0.5cm}
\end{figure}


Figure~\ref{fig:developer} shows the distribution of years of work experience of our participants. The average number of years of  work experience that these participants have is 3.5.
Two developers from the educational institution are very senior, who have worked for 5.5 and 10 years, respectively.
The most experienced developer from industry has worked for seven years, while some has only worked for one year. Based on their working experience, we group participants into two groups: \emph{junior} and \emph{senior}. There are 20 \emph{junior} developers and 15 \emph{senior} developers, respectively.


We divided the 35 participants into seven groups. The ratio of \emph{junior} and \emph{senior} developers for each group was kept approximately the same. Each patch generated by program repair tools is labeled by five participants. Participants in the same group receive the same set of patches to label.

\section{Assessing Independent Annotators' Labels}\label{sec:rq1}


Our user study presented in Section~\ref{sec:setup} was conducted to build a set of gold standard labels for machine-generated patches, which can reliably be used to assess reliability of author and automated annotations. Before using the labels produced by our user study, we need to first ascertain their quality. Agreement among annotators is often used as a measure of quality~\cite{christopher2008introduction,DamessieNSC17,ScholerTS11}. Thus, in this section, we investigate the degree to which the annotators agree with one another. This answers RQ1: Can independent annotators agree on patch correctness?

\vspace{0.2cm}\noindent{\bf Methodology.} To answer RQ1, we first compute some simple statistics highlighting the number of agreements and disagreements among annotators. We then calculate several well-accepted measures of inter-rater reliability. Finally, we perform some sanity checks to substantiate whether or not annotators are arbitrary in making their decisions.

\vspace{0.2cm}\noindent{\bf Results.} To recap, our annotators are 35 professional developers who are tasked to annotate 189 machine-generated patches. Each patch is annotated by five professional developers; each provides either one of the following labels: incorrect, correct, or unknown. Table~\ref{tbl:devlabels} summarizes the number of agreements and disagreements among annotators. The number of patches in which all developers agree on each patch's label is 118 (62.4\% of all patches); of which 95 patches are labeled as incorrect and 23 patches are labeled as correct. Moreover, ignoring unknown labels, the number of patches for which the remaining annotators fully agree on their labels is 155 (82.0\% of all patches). Out of these, the numbers of patches that are labeled as incorrect and correct are 132 and 23, respectively. Lastly, for 187 out of 189 patches (98.9\% of all patches), there is a majority decision (i.e., most annotators agree on one label). Out of these, 152 and 35 patches are identified as incorrect and correct, respectively.


\begin{table}
\caption{Results of participant annotations. First column indicates the number of patches that every developer agrees on the label of each patch as correct or incorrect. Second column indicates the number of patches, wherein each patch has least one developer labeling it as unknown and the remaining developers agrees on the label of the patch. Last column indicates the number of patches that the label of each patch can be determined by a majority voting among developers' labels. \label{tbl:devlabels}}
\vspace{-0.3cm}
\begin{center}
{\small
  \def\arraystretch{0.8}
 \begin{tabular}{l|rrr}
 \toprule
          &\textbf{All Agree} &\textbf{All Agree - Unk}&
          \textbf{Majority Agree}\\

 \midrule
  Incorrect      & 95 & 132 & 152\\
  Correct        & 23 & 23 & 35\\
  \midrule
  Total        & 118 & 155 & 187\\
 \bottomrule
 \end{tabular}}
 \end{center}
 \vspace{-0.5cm}
 \end{table}

We also compute several inter-rater reliability scores: mean pairwise Cohen's kappa~\cite{christopher2008introduction,cohen1960coefficient} and Krippendorff's alpha~\cite{Krippendorff70}. Using the earlier test we consider three different ratings (i.e., correct, incorrect, and unknown), while the latter test allows us to ignore unknown ratings\footnote{Krippendorff's alpha allows us to have different number of ratings for each data point.}. Inter-rater reliability scores measure how much homogeneity, or consensus, there is between raters/labelers. The importance of rater reliability hinges on the fact that it represents the extent to which the data collected in the study are correct representations of the variables being measured. A low inter-rater reliability suggests that either the rating scale used in the study is defective or raters need to be retrained for the rating task or the task is highly subjective. The higher the inter-rater reliability the more reliable the data is.

\begin{table}
\centering
\caption{Interpretation of Inter-Rater Reliability Scores by Landis and Koch~\cite{landis1977measurement}.}\vspace{-0.2cm}
\label{tbl:kappalandis}
  {\small
   \def\arraystretch{0.85}
\begin{tabular}{|l|l|}
\hline
\textbf{Score Range}              & \textbf{Interpretation}                     \\          \hline
$<0$            & poor agreement                                                   \\ \hline
$[0.01, 0.20]$            & slight agreement                                                   \\ \hline
$[0.21, 0.40]$            & fair agreement                                                   \\ \hline
$[0.41, 0.60]$            & moderate agreement                                                   \\ \hline
$[0.61, 0.80]$            & substantial agreement                                                   \\ \hline
$[0.81, 1.00]$            & almost perfect agreement                                                   \\ \hline
\end{tabular}}
\end{table}


Table~\ref{tbl:kappalandis} shows details of interpretations of reliability score values by Landis and Koch ~\cite{landis1977measurement}. It is worth noting that there is another interpretation of kappa value by Manning \textit{et al.}~\cite{christopher2008introduction}, which indicates that a kappa value falling between 0.67 and 0.8 demonstrates a fair agreement between raters -- the second highest level of agreement by their interpretation. It has been shown that this fair level of inter-rater agreement normally happens in popular datasets such as those used for TREC evaluations\footnote{Text REtrieval Conference (TREC), which is championed by US National Institute of Standards and Technology (NIST) since 1992, provides benchmark datasets for various text retrieval tasks -- see \url{http://trec.nist.gov/data.html}.}  and medical IR collections~\cite{christopher2008introduction}.


\vspace{0.1cm}\begin{tcolorbox}[width=8.5cm]
The computed mean pairwise Cohen's kappa and Krippendorff's alpha for our data are 0.691 and 0.734 respectively, which highlight a substantial agreement among participants and satisfies the standard normally met by quality benchmark datasets.\vspace{0.1cm}
\end{tcolorbox}

\begin{figure}
\centering
\includegraphics[width=0.3\textwidth,draft=false]{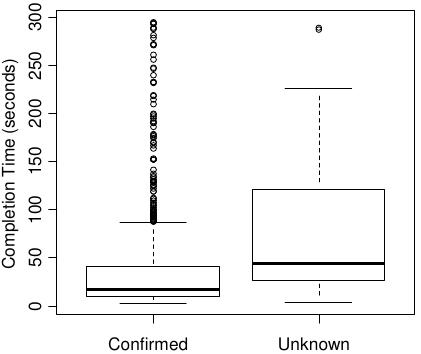}\vspace{-0.2cm}
 \caption{Time taken by annotators to decide whether a patch's label is either known (confirmed as correct or incorrect) or unknown.}
 \label{fig:hypothesis1}
 \vspace{-0.5cm}
\end{figure}

\begin{figure}[!t]\vspace{-0.3cm}
\centering
\includegraphics[width=0.3\textwidth,draft=false]{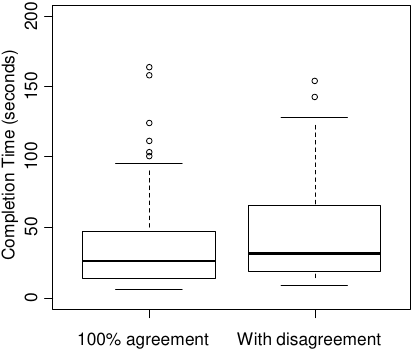}\vspace{-0.2cm}
 \caption{Time taken by annotators to decide a patch's label for full-agreement and disagreement cases.}
 \label{fig:hypothesis2}
 \vspace{-0.4cm}
\end{figure}

To further validate the annotations, we perform two sanity checks to substantiate whether or not annotators are arbitrary in their decisions:

\begin{itemize}[leftmargin=*]

\item First, we expect conscientious annotators to spend more time inspecting patches that are eventually labeled as unknown than other patches. Annotators who label patches as unknown without thinking much would be likely making arbitrary decisions. Figure~\ref{fig:hypothesis1} depicts a box plot showing the time participants took on patches that are labeled as unknown and other patches. It can be seen that participants took more time on the earlier set of patches. Wilcoxon signed-rank test returns a p-value that is less than 0.005, indicating a statistically significant difference. Moreover, the Cliff's delta\footnote{Cliff defines a delta of less than 0.147, between 0.147 to 0.33, between 0.33 and 0.474, and above 0.474 as negligible, small, medium, and large effect size, respectively~\cite{Cliff93}.}, which is a non-parametric effect size measure, is 0.469 (medium).

\item Second, we expect conscientious annotators to spend more time inspecting difficult patches than easy ones. We consider disagreement among annotators as proxy for patch difficulty. We compare the time taken by participants in identifying patches for which there is complete agreement to those for which disagreement exists. Figure~\ref{fig:hypothesis2} shows a box plot which shows that participants spend more time on disagreement cases. Wilcoxon signed-rank test returns a p-value that is less than 0.05, indicating statistically significant difference. Moreover, the Cliff's delta is 0.178 (small).

\end{itemize}

The above results substantiate the quality of our dataset. In the subsequent sections, which answer RQ2 and RQ3, we use two versions of our dataset ALL-AGREE (see ``All Agree'' column in Table~\ref{tbl:devlabels}) and MAJORITY-AGREE (see ``Majority Agree'' column in Table~\ref{tbl:devlabels}), to assess the reliability of author and automated annotations.


%

\section{Assessing Author Annotation}\label{sec:assessauthor}

A number of studies proposing automated repair approaches evaluate the proposed approaches through manual annotation performed by authors~\cite{liu2017identifying,xiong2017precise,LeLG16}. Author subjectivity may cause bias which can be a threat to the internal validity of the study. Author bias has been actively discussed especially in the medical domain, e.g.,~\cite{Vaccaro11}. Unfortunately so far, there has been no study that investigates presence or absence of bias in author annotation and its impact to the validity of the labels in automated program repair. This section describes our effort to fill this need by answering RQ2: How reliable is author annotation?

\vspace{0.1cm}\noindent{\bf Methodology.} Recall that our user study makes use of patches released by three research groups, including Liu et al.~\cite{liu2017identifying}, Martinez et al.~\cite{DBLP:journals/ese/MartinezDSXM17}, and Le et al.~\cite{le2017s3} who created program repair tools namely ACS, Nopol, and S3, respectively. Authors of each tool manually labeled the patches generated by their tool and its competing approaches by themselves. To answer RQ2, we compare labels produced by the three research groups with those produced by our independent annotators whose quality we have validated in Section~\ref{sec:rq1}. We consider the ALL-AGREE and MAJORITY-AGREE datasets mentioned in Section~\ref{sec:rq1}.


\begin{table}
\caption{Results of labels by authors compared to independent annotators. \label{tbl:rq2truthathor}}
\vspace{-0.3cm}
\begin{center}
{\small
 \def\arraystretch{0.7}
 \begin{tabular}{ll|rr}
 \toprule
         & Indep Annotators-Authors &\textbf{All Agree} &
          \textbf{Majority Agree}\\

 \midrule
 \multirow{2}{*}{Same}
 & Incorrect-Incorrect      & 82 & 133 \\
 & Correct-Correct        & 23 & 33 \\
  \midrule
 \multirow{4}{*}{Different}
 & Incorrect-Correct      & 6 & 10 \\
 & Correct-Incorrect        & 0 & 2 \\
 & Incorrect-Unknown        & 7 & 9 \\
 & Correct-Unknown        & 0 & 0 \\
  \midrule
  \multicolumn{2}{c|}{Total} & 118 & 187\\
 \bottomrule
 \end{tabular}}
 \end{center}
 \vspace{-0.1cm}
 \end{table}


\vspace{0.1cm}\noindent{\bf Results.} Table~\ref{tbl:rq2truthathor} shows the detailed results on the comparisons between authors' labels and independent annotators' labels. We found that for ALL-AGREE dataset, authors' labels match with independent annotators' labels (Same) for 105 out of 118 patches (89.0\%). There are 13 patches for which authors' labels mismatch those by independent annotators (Different). Among these patches, 6 are identified by independent annotators as incorrect, but identified by authors as correct (Incorrect-Correct). For the other 7 patches, authors' labels are unknown while independent annotators' labels are incorrect (Incorrect-Unknown). For the MAJORITY-AGREE dataset, 88.8\% of the labels match. There are 21 mismatches; 10 belong to Incorrect-Correct cases, 2 to Correct-Incorrect cases, and 9 to Incorrect-Unknown cases. Figure~\ref{fig:rq2ex1} shows an example patch generated by Nopol~\cite{nopol:xuan} that has mismatched labels. It is labeled as correct by Martinez et al. and incorrect by independent annotators.

\lstset{language=Java}
\definecolor{dkgreen}{rgb}{0,0.5,0}
\definecolor{dkred}{rgb}{0.5,0,0}
\definecolor{gray}{rgb}{0.5,0.5,0.5}
\lstset{
basicstyle=\ttfamily\bfseries\scriptsize,
  morekeywords={virtualinvoke},
  keywordstyle=\color{blue},
  ndkeywordstyle=\color{red},
  commentstyle=\color{dkred},
  stringstyle=\color{dkgreen},
  numbers=left,
  breaklines=true,
  numberstyle=\ttfamily\footnotesize\color{gray},
  stepnumber=1,
  numbersep=10pt,
  backgroundcolor=\color{white},
  tabsize=4,
  showspaces=false,
  showstringspaces=false,
  xleftmargin=.23in
}
\begin{figure}
\begin{lstlisting}[frame=none,mathescape=true]
@@ -115,9 +115,7 @@ public class StopWatch {
public void stop() {
   if(this.runningState != STATE_RUNNING && this.runningState != STATE_SUSPENDED) {
       throw new IllegalStateException("...");
   }
+  if(this.runningState == STATE_RUNNING)// Developer patch
+  if(-1 == stopTime)// Generated patch
       stopTime = System.currentTimeMillis();
   this.runningState = STATE_STOPPED;
}
\end{lstlisting}
\vspace{-0.4cm}
\caption{An example of a patch that has mismatched labels. Martinez et al. identified the patch (at line 7) as correct, while independent annotators identified this patch as incorrect. The ground truth (developer) patch is shown at line 6.}
\label{fig:rq2ex1}
\vspace{-0.5cm}
\end{figure}



We also compute inter-rater reliability of authors' labels and labels in ALL-AGREE and MAJORITY-AGREE datasets. The Cohen's kappa values are 0.719 and 0.697 considering the ALL-AGREE and MAJORITY-AGREE datasets, respectively\footnote{The Krippendorf's alpha values are 0.717 and 0.695}. Comparing these scores with Landis and Koch's interpretation in Table~\ref{tbl:kappalandis}, there is substantial agreement. 



\vspace{0.1cm}\begin{tcolorbox}[width=8.5cm]
A majority (88.8-89.0\%) of patch correctness labels produced by author annotation match those produced by independent annotators. Inter-rater reliability scores indicate a substantial agreement between author and independent annotator labels.
\end{tcolorbox}





\begin{figure}[t!]\vspace{-0.3cm}
\centering
\includegraphics[width=0.3\textwidth,draft=false]{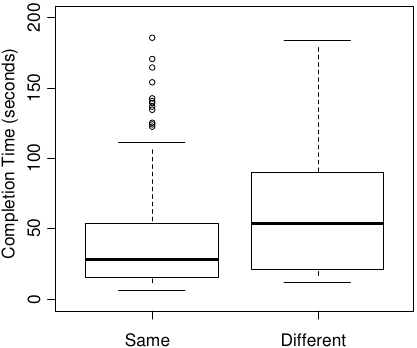}
\vspace{-0.3cm}
 \caption{Participant completion time for patches for which author and independent annotator labels match (Same) and those whose labels mismatch (Different)}
 \label{fig:rq2majoragree}
 \vspace{-0.68cm}
\end{figure}

To better characterize cases where author and independent annotator labels match (Same) and those where they do not match (Different), we investigate the time that participants of our user study took to label the two sets of patches. Since the number of mismatches is smaller in the ALL-AGREE dataset, we focus on comparing labels in MAJORITY-AGREE dataset. Figure~\ref{fig:rq2majoragree} depicts a box plot showing the distribution of completion time corresponding to the two sets of patches. According to the figure, patches with matching labels took participants a shorter period of time to label comparing to those whose labels mismatched. Wilcoxon signed-rank test returns a p-value that is less than 0.05, indicating statistically significant difference. The Cliff's delta is equal to 0.278 (small). Since task completion time can be used as a proxy for measuring task difficulty or lack thereof~\cite{wickens1991processing}, we consider participants completion time as a proxy of difficulty in assessing patch correctness. The result suggests that disagreements between authors and independent annotators happen for more difficult cases.

\section{Assessing Automated Annotation}\label{sec:assessautomated}

In this research question, we investigate the reliability of the use of automatically generated independent test suite (ITS) in annotating patch labels. ITS has been used as an objective proxy to measure patch correctness -- a patch is deemed as incorrect if it does not pass the ITS, and as correct or generalizable otherwise~\cite{smith2015cure,le2016empirical}. It is unequivocal that incorrect patches determined by ITS are indeed incorrect. However, it is unclear if ITS can detect a large proportion of incorrect patches. Moreover, the extent to whether correct (generalizable) patches determined by ITS are indeed correct remain questionable. Thus, to assess the usefulness of ITS, we investigate the answer to RQ3: How reliable is automatically generated ITS in determining patch correctness?



\vspace{0.1cm}
\noindent\textbf{Methodology:} We employ the recently proposed test case generation tool \textsc{DiffTGen} by Xin \textit{et. al}~\cite{xin2017identifying} and \textsc{Randoop}~\cite{DBLP:conf/icse/PachecoLEB07} to generate ITS. To generate ITS using \textsc{DiffTGen}  and \textsc{Randoop}, the human-patched program is used as ground truth. For \textsc{DiffTGen}, we run using its best configuration reported in~\cite{xin2017identifying}, allowing it to invoke \textsc{Evosuite}~\cite{DBLP:conf/sigsoft/FraserA11} in 30 trials with the search time of each trial limited to 60 seconds. A machine-generated patch is identified as \emph{incorrect} if there is a test in the \textsc{DiffTGen}-generated ITS that witnesses the output differences between the machine and human patches. For \textsc{Randoop}, we run it on the ground truth program with 30 different seeds with each run limited to 5 minutes. A machine-generated patch is identified as incorrect if there is at least one test case in the \textsc{Randoop}-generated ITS that exhibits different test results in machine-patched and human-patched (ground truth) programs, e.g., it fails on the machine-patched program but passes on the ground truth program, or otherwise. By this way, we allow both tools to generate multiple test suites. It is, however, worth noting that \textsc{DiffTGen} and \textsc{Randoop} are incomplete in the sense that they do not guarantee to always generate the test cases that witness incorrect patches.


 We use test cases generated by the tools to automatically annotate the 189 patches and compare the generated labels to those in ALL-AGREE and MAJORITY-AGREE datasets which are created by our user study.

\vspace{0.1cm}
\noindent\textbf{Results:} Out of the 189 patches in our study, \textsc{DiffTGen} generates test cases that witness 27 incorrect (overfitting) patches. Details of these patches are shown in Table~\ref{tbl:rq3}. The ALL-AGREE ground truth identifies 17 of these 27 patches as incorrect (the other 10 patches lie outside of the ALL-AGREE dataset), while the MAJORITY-AGREE dataset identifies all of them as incorrect. Unfortunately, most of the patches labelled as incorrect in ALL-AGREE (65 patches) and MAJORITY-AGREE (121 patches) datasets failed to be detected as such by ITS generated by \textsc{DiffTGen}. \textsc{Randoop} performs similarly as compared to \textsc{DiffTGen}. It identifies 31 patches as incorrect, all of which are also identified as incorrect in the MAJORITY-AGREE dataset. Note that, \textsc{DiffTGen} and \textsc{Randoop} when combined can identify totally 51 unique patches as incorrect.

In their studies, Smith et al.~\cite{smith2015cure} and Le et al.~\cite{smith2015cure} assume a patch is incorrect if it does not pass an ITS, and correct or generalizable otherwise. Using the same assumption to generate correctness labels, we can compute inter-rater reliability between labels automatically annotated by running ITS generated by \textsc{DiffTGen} and \textsc{Randoop} and labels in ALL-AGREE and MAJORITY-AGREE datasets. As readers may have expected, the kappa values are very low as shown in Table~\ref{tbl:kappaRQ3}, e.g., Cohen's kappa values when using \textsc{DiffTGen}-generated ITS for ALL-AGREE and MAJORITY-AGREE are  0.078 and 0.075, repsectively.\footnote{The corresponding Krippendorff's alpha values are -0.32 and -0.336}

\begin{table}\vspace{-0.3cm}
\caption{Kappa values when using \textsc{DiffTGen}, \textsc{Randoop}, and their combination to label patches in ALL-AGREE and MAJORITY-AGREE datasets.}
\label{tbl:kappaRQ3}
\vspace{-0.3cm}
\begin{center}
{\small
  \tabcolsep 2.5pt
   \def\arraystretch{0.8}
 \begin{tabular}{lrrr|rrr}
 \toprule
          &\multicolumn{3}{c}{\textbf{All Agree}}  &\multicolumn{3}{c}{\textbf{Majority Agree}}\\
          \cline{2-7}
          & \textsc{DiffT} & \textsc{Rand} & \textsc{Comb}
 & \textsc{DiffT} & \textsc{Rand} & \textsc{Comb} \\
  \midrule
  Cohen's Kappa      & 0.078 & 0.073 & 0.158  & 0.075 & 0.072 & 0.146\\
  Kripp's Alpha        & -0.32 & -0.3 & -0.057 & -0.336 &-0.313 &-0.097\\
 \bottomrule
 \end{tabular}}
 \end{center}
 \vspace{-0.5cm}
 \end{table}

\vspace{0.1cm}\begin{tcolorbox}[width=8.5cm]
Independent test suite generated by \textsc{DiffTGen} and \textsc{Randoop} can only label fewer than a fifth of incorrect patches as such in ALL-AGREE and MAJORITY-AGREE datasets.
\end{tcolorbox}




We now compare author labels discussed in Section~\ref{sec:assessauthor} with ITS labels. Table~\ref{tbl:rq3} shows the author labels of the 27 and 31 patches identified as incorrect by \textsc{DiffTGen} and \textsc{Randoop}, respectively. For these patches, the majority of the labels by authors and \textsc{DiffTGen} match. However, there are three special patches identified as incorrect by \textsc{DiffTGen}, including Math\_80 generated by Kali, Chart\_3 generated by GenProg, and Math\_80\_2015 generated by Nopol, while author labels are ``Unknown''. One special patch identified as incorrect by \textsc{Randoop} (Math\_73 generated by GenProg), is labelled as correct by authors.


\begin{table} [!htb]\vspace{-0.3cm}
 \caption{Labels by Independent annotators (``Annot'' column) and authors (``Authors'' column) of patches identified by independent test suite (ITS) generated by \textsc{DiffTGen} or \textsc{Randoop} as incorrect .} \label{tbl:rq3}
 \vspace{-10pt}
 \begin{center}
 {\small
  \tabcolsep 2.5pt
  \def\arraystretch{0.80}
 \begin{tabular}{ll|rrrr}
 \toprule
          & & \textbf{\textsc{DiffTGen}} & \textbf{\textsc{Randoop}} &\textbf{Annot} &
          \textbf{Authors} \\

 \midrule
  \multirow{10}{*}{Kali}  & Time\_4 & Incorrect & Incorrect & Incorrect & Incorrect  \\
   & Math\_32 & Incorrect &  & Incorrect& Incorrect  \\
    & Math\_2  & Incorrect &  & Incorrect & Incorrect   \\
     & Math\_80  & Incorrect &  & Incorrect & \underline{Unknown}  \\
      & Math\_95  & Incorrect & Incorrect & Incorrect & Incorrect \\
       & Math\_40   & Incorrect &  & Incorrect & Incorrect \\
        & Chart\_13  & Incorrect &  & Incorrect & Incorrect  \\
         & Chart\_26  & Incorrect &  & Incorrect & Incorrect   \\
          & Chart\_15  & Incorrect & Incorrect & Incorrect & Incorrect \\
           & Chart\_5   & Incorrect & Incorrect & Incorrect & Incorrect \\
  \midrule
  \multirow{10}{*}{GenProg}    & Math\_2  & Incorrect & & Incorrect & Incorrect\\
  & Math\_8  & Incorrect & & Incorrect & Incorrect \\
       & Math\_80   & Incorrect & & Incorrect & Incorrect  \\
       & Math\_81   & Incorrect & & Incorrect & Incorrect \\
        & Math\_95   & Incorrect & Incorrect & Incorrect & Incorrect \\
         & Math\_40   & Incorrect & & Incorrect & Incorrect \\
         & Math\_73   &  & Incorrect & Incorrect & \underline{Correct} \\
         & Chart\_1   & & Incorrect & Incorrect & Incorrect \\
          & Chart\_3   & Incorrect & & Incorrect & \underline{Unknown} \\
           & Chart\_5   & Incorrect & Incorrect & Incorrect & Incorrect  \\
           & Chart\_15   & Incorrect & Incorrect & Incorrect & Incorrect  \\
  \midrule
  \multirow{27}{*}{Nopol}    & Math\_33   & Incorrect & & Incorrect & Incorrect \\
   & Math\_73\_2017   & & Incorrect & Incorrect & Incorrect \\
        & Math\_80\_2017   & Incorrect & & Incorrect & Incorrect  \\
         & Math\_80\_2015   & Incorrect & & Incorrect & \underline{Unknown}  \\
         & Math\_97   & Incorrect & & Incorrect & Incorrect \\
         & Math\_105   & & Incorrect & Incorrect & Incorrect \\
         & Time\_16   & & Incorrect & Incorrect & Incorrect \\
          & Time\_18   & & Incorrect & Incorrect & Incorrect \\
           & Chart\_13\_2017   & Incorrect & & Incorrect & Incorrect \\
            & Chart\_13\_2015   & Incorrect & & Incorrect & Incorrect \\
            & Chart\_21\_2017   & Incorrect & & Incorrect & Incorrect \\
            & Chart\_21\_2015   & Incorrect & & Incorrect & Incorrect  \\
            & Closure\_7   & & Incorrect & Incorrect & Incorrect \\
              & Closure\_12   & & Incorrect & Incorrect & Incorrect \\
                            & Closure\_14   & & Incorrect & Incorrect & Incorrect \\
            & Closure\_20   & & Incorrect & Incorrect & Incorrect \\
            & Closure\_30   & & Incorrect & Incorrect & Incorrect \\
                        & Closure\_33   & & Incorrect & Incorrect & Incorrect \\
                        & Closure\_76   & & Incorrect & Incorrect & Incorrect \\
                                      & Closure\_111   & & Incorrect & Incorrect & Incorrect \\
                        & Closure\_115   & & Incorrect & Incorrect & Incorrect \\
                          & Closure\_116   & & Incorrect & Incorrect & Incorrect \\
                          & Closure\_120   & & Incorrect & Incorrect & Incorrect \\
                                        & Closure\_124   & & Incorrect & Incorrect & Incorrect \\
                           & Closure\_130   & & Incorrect & Incorrect & Incorrect \\
                           & Closure\_121   & & Incorrect & Incorrect & Incorrect \\
                                       & Mockito\_38   & & Incorrect & Incorrect & Incorrect \\
    \midrule
    \multirow{1}{*}{Angelix}    & Lang\_30   & & Incorrect & Incorrect & Incorrect \\
     \midrule
     \multirow{1}{*}{CVC4}    & Lang\_30   & & Incorrect & Incorrect & Incorrect \\
         \midrule
         \multirow{1}{*}{Enum}    & Lang\_30   & & Incorrect & Incorrect & Incorrect \\
 \bottomrule
 \end{tabular}}
 \end{center}
 \vspace{-0.5cm}
 \end{table}

\begin{figure}[t]
\centering
\includegraphics[width=0.3\textwidth,draft=false]{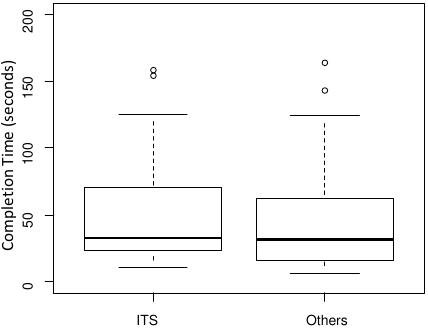}\vspace{-0.3cm}
\caption{Participant completion time for the 51 unique patches labelled by \textsc{DiffTGen}'s and \textsc{Randoop}'s ITSs as incorrect versus that for other patches.}
\label{fig:rq3itsothers}
\vspace{-0.5cm}
\end{figure}

Finally, we want to investigate the difficulty of judging correctness of patches that \textsc{DiffTGen} and \textsc{Randoop} generated ITSs label as incorrect. To do so, we compare participant completion time for the set of 51 unique patches and the set of other patches. Figure~\ref{fig:rq3itsothers} shows time spent by participants labelling these two sets of patches. We find that they are more or less the same. Wilcoxon signed-rank test confirms that the difference is not statistically significant. Thus, patches that ITS successfully label as incorrect are not necessarily the ones that participants require more time to manually label.




\section{Discussion}\label{sec:discussion}
In this section, we first provide implications of our findings. We then discuss our post-study survey, in which we asked a number of independent annotators for rationales behind their patch correctness judgements. At the end of this section, we discuss some threats to validity.

\subsection{Implications}\label{subsec:implication}
To recap, we have gained insights into the reliability of patch correctness assessment by authors and by automatically generated independent test suite (ITS); each of them has their own advantages and disadvantages. Based on these insights, we provide several implications as follows.

\begin{tcolorbox}[width=8.5cm]
Authors' evaluation of patch correctness should be made publicly available to the community.
\end{tcolorbox}

Liu et al., Martinez et al., and Le et al. released their patch correctness labels publicly~\cite{liu2017identifying, DBLP:journals/ese/MartinezDSXM17, le2017s3}, which we are grateful for. We believe that considerable effort has been made by authors to ensure the quality of the labels. Still, we notice that for slightly more than 10\% of the patches, authors' labels are different from the ones produced by multiple independent annotators. Thus, we encourage future ASR paper authors to release their datasets for public inspection. The public (including independent annotators) can then provide inputs on the labels and possibly update labels that may have been incorrectly assigned. Our findings here (e.g., author annotations are fairly reliable) may not generalize to patches labelled by authors which have not been released publicly. It is possible that the quality of correctness labels for those patches (which are not made publicly available) to be lower.  Also, as criticized by Monperrus \textit{et al.}~\cite{monperrus2014critical}, the conclusiveness of the evaluation of techniques that keep patches and their correctness labels private is questionable.



\vspace{0.1cm}\begin{tcolorbox}[width=8.5cm]
Collaborative effort is needed to distribute the expensive cost of ASR evaluation. 
\end{tcolorbox}


In this study, we have evaluated correctness of 189 automatically generated patches by involving independent annotators. We have shown that the quality of the resultant labels (measured using inter-rater reliability) are on par with high-quality text retrieval benchmarks~\cite{christopher2008introduction}. Unfortunately, evaluation using independent annotators is expensive. To evaluate 189 patches, we need to get 35 professional developers; Each agrees to spend up to an hour of their time. This process may not be scalable especially considering the large number of new ASR techniques that are released in the literature year by year. Thus, there is a need for a more collaborative effort to distribute the cost of ASR evaluation. One possibility is to organize a competition involving impartial industrial data owners (e.g., software development houses willing to share some of their closed bugs) who are willing to judge correctness of generated patches. Similar competitions with industrial data owners have been held to advance various fields such as forecasting\footnote{\url{http://www.cikm2017.org/CIKM_AnalytiCup_task1.html}} and fraud detection\footnote{\url{http://research.larc.smu.edu.sg/fdma2012/}}.


\begin{figure}\vspace{-0.3cm}
\centering
\includegraphics[width=0.4\textwidth,draft=false]{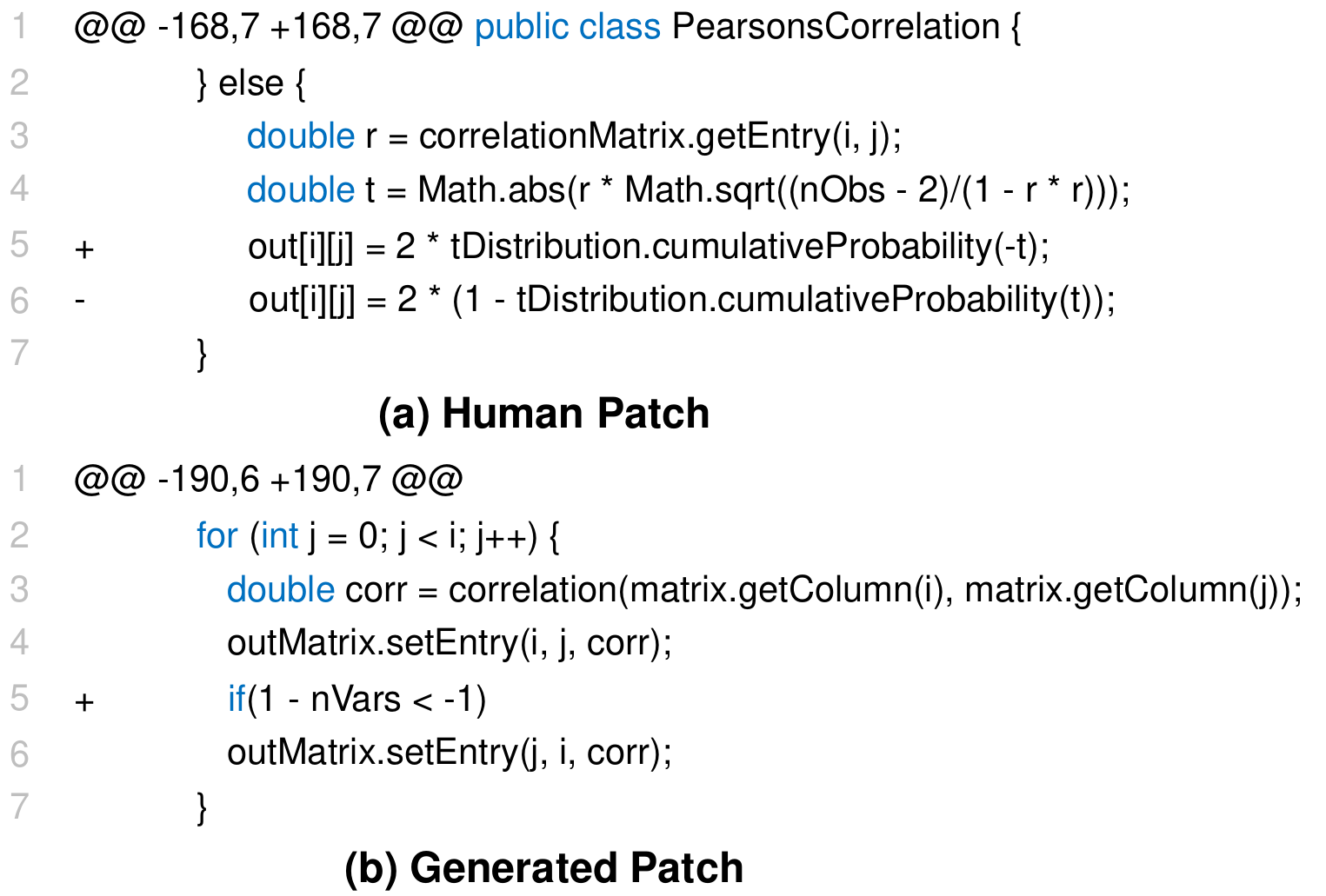}\vspace{-0.3cm}
 \caption{A machine-generated patch labeled by ITS as incorrect but labeled by author annotation as unknown.}
 \label{fig:implexample}
 \vspace{-0.5cm}
\end{figure}

\vspace{0.1cm}\begin{tcolorbox}[width=8.5cm]
Independent test suite (ITS) {\em alone} should not be used to evaluate the effectiveness of ASR.
\end{tcolorbox}

Independent test suites (ITSs) generated by \textsc{DiffTGen}~\cite{xin2017identifying} and \textsc{Randoop}~\cite{DBLP:conf/icse/PachecoLEB07} have been shown to be ineffective in annotating correctness labels for patches (see Section~\ref{sec:assessautomated}). Only fewer than a fifth of the incorrect patches are identified as such by ITSs generated by \textsc{DiffTGen} and \textsc{Randoop}. Based on effectiveness of state-of-the-art test generation tool for automatic repair that we assessed in this study, we believe that ITS {\em alone} should not be used for {\em fully automated} patch labeling. The subject of ITS generation for program repair is new though and we encourage future studies to improve the quality of automatic test generation tools so that more incorrect patches can be detected. That being said, automated patch annotation may not be a silver bullet; the general problem of patch correctness assessment (judging the equivalence of developer patch and automatically generated patch) is a variant of program equivalence problem which has been proven to be undecidable with no algorithmic solution~\cite{Sipser:1996:ITC:524279}.

\vspace{0.1cm}\begin{tcolorbox}[width=8.5cm]
Independent test suite, despite being less effective, can be used to augment author annotation.
\end{tcolorbox}

It has been shown in Section~\ref{sec:assessautomated} that ITS generated by \textsc{DiffTGen} and \textsc{Randoop} identified four patches as incorrect whereas the labels generated by author annotation are unknown and correct. An example of such patch is shown in Figure~\ref{fig:implexample}. From the figure, we can notice that it is hard to judge whether the patch is correct or incorrect. From this finding, we believe that ITS, despite being less effective than author annotation in identifying correct patches, can be used to augment author annotation by helping to resolve at least some of the ambiguous cases. Authors can possibly run \textsc{DiffTGen} and \textsc{Randoop} to identify clear cases of incorrect patches; the remaining cases can then be manually judged. The use of both author and automated annotation via ITS generation can more closely approximate multiple independent annotators' labels while requiring less cost.



\subsection{Post-Study Survey}
We conducted a post-study survey to investigate why a developer chooses a different answer from the majority.
Among the 189 patches, there are several patches where the majority, but not all participants, agree on patch correctness. Among participants annotating these patches, we selected 11 who answered differently from the majority and emailed them to get deeper insights into their judgments. In our email, we provided a link to the same web interface used in our user study to allow participants to revisit their decision for the patch in question. Notice that we did not inform the participants that their answers were different from the majority. We received replies from 8 out of the 11 participants (72.7\% response rate).

\begin{figure}[t]\vspace{-0.3cm}
\centering
\includegraphics[width=0.35\textwidth,draft=false]{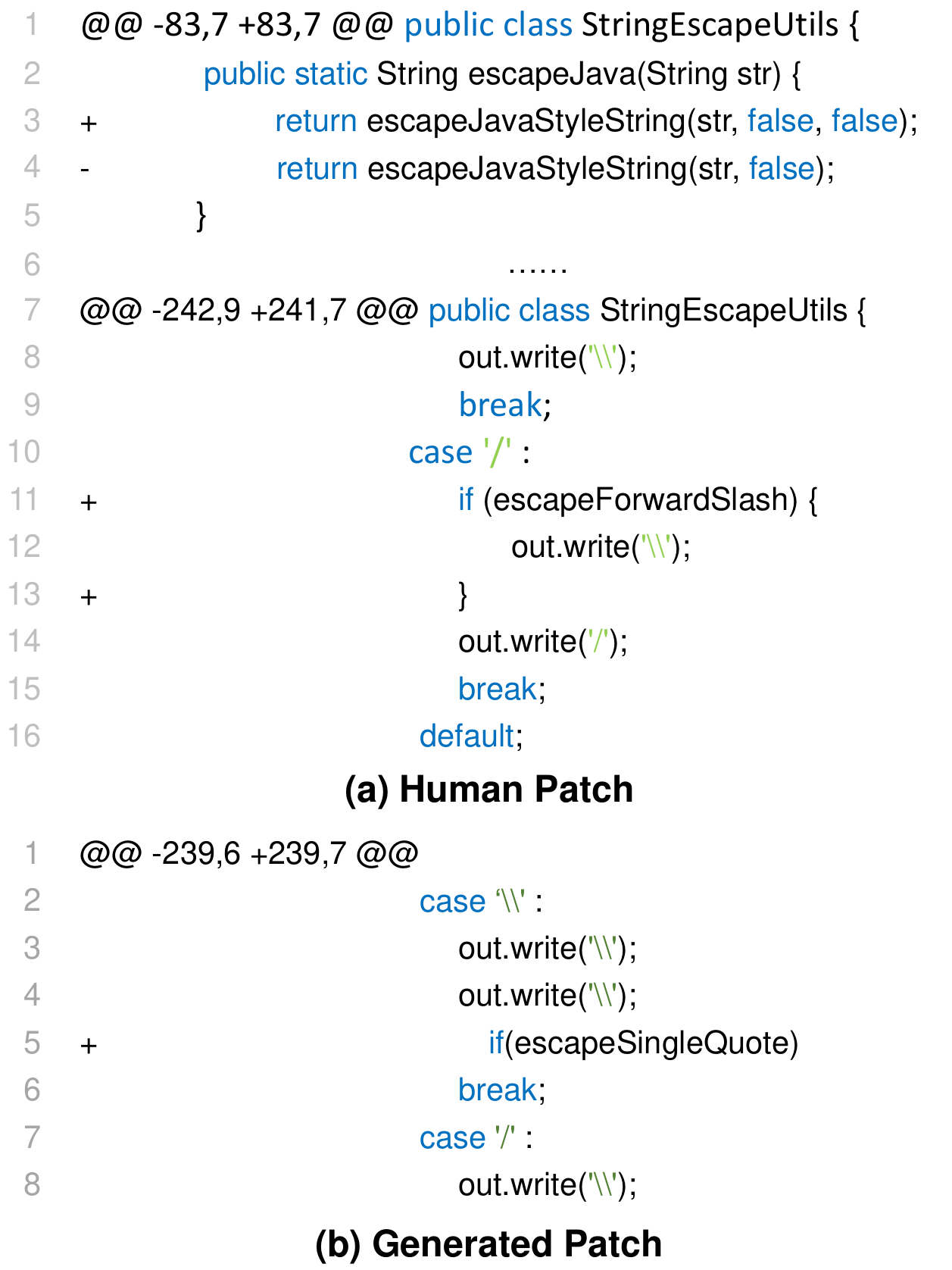}\vspace{-0.3cm}
 \caption{An example of a patch in post-study}
 \label{fig:poststudy}
 \vspace{-0.5cm}
\end{figure}

We found that 5 out of 8 developers changed their correctness labels after they looked into the patch again; their revised labels thus became consistent with the labels that the majority agree. The remaining three kept their correctness labels; two judged two different patches as incorrect (while the majority labels are correct) while another judged a patch as correct (while the majority label is incorrect). These participants kept their decision for different reasons; one was unsure of a complex expression involved in the patch, another highlighted a minor difference that may be considered ignorable by others, and the other participant viewed the generated and ground truth patch to have similar intentions. An excerpt of the patch in question for the last mentioned participant is shown in Figure~\ref{fig:poststudy}.

\subsection{Threats to Validity}
\noindent{\bf Threats to internal validity.} These threats relate to potential errors and biases in our study. The following are a few relevant threats that deserve further discussion:

\begin{itemize}[leftmargin=*]
\item There may be errors in the web interface, that we provide to participants in our study, and the code for analyzing the collected data. To reduce the possibility of errors in the web interface, we conducted a pilot study with a few graduate students and incorporated their feedback. We also performed a thorough check on our code.

\item Due to constrained resources (we only have 35 professional developers agreeing to devote an hour of their time), we do not use all patches in the original dataset by Liu et al.~\cite{liu2017identifying}, Martinez et al.~\cite{DBLP:journals/ese/MartinezDSXM17}, and Le et al.~\cite{le2017s3}. If the whole collection of patches is used, it is possible that results may differ. To mitigate this threat, we randomly selected patches that are included in this study while keeping the ratios of patches coming from different ASR tools approximately the same.

\item Professional developers included in our user study are not the original developers of the buggy code and ground truth patches. Unfortunately, since the original developer patches included in Liu et al.'s study were committed many years ago (the earliest being 2006), it is hard to get the original developers to participate in our study. Also, the original developers may also have forgotten the detail of the patches. Additionally, since the patches are small, the task of comparing two patches and judging whether they are equivalent or not should be managable to professional developers. Indeed, in our study, our respondents are able to provide definite labels to a majority of patches (i.e., only 44 out of 750 labels (5.9\%) are unknown, while the rest are either incorrect or correct). To improve the reliability of the labels, we ask not only one professional developer but five of them. As highlighted in Section~\ref{sec:rq1} there is a substantial agreement among participants satisfying standard followed by high-quality benchmark datasets. Furthermore, to help developers understand patches, we also provide multiple resources including source code files, failed test cases, {\sc GitHub} link of the project, etc. A large number of past software engineering, studies e.g.,~\cite{gachechiladze2017anger, buse2010learning, baysal2014no, ko2014thirty, gachechiladze2017anger, daka2015modeling, ormandjieva2007toward} have also involved third-party labellers (who are not content creators) to assign labels to data. The same situation was also followed in other related areas, e.g., information retrieval~\cite{damessie2017gauging,bailey2008relevance}. We also make the 189 patches and participants' responses publicly available for public inspection.\footnote{URL omitted for double blind reviewing but would be made available later.}

\end{itemize}


\noindent{\bf Threats to external validity.} These threats relate to the generalizability of our results. The following are a few relevant threats that deserve further discussion:

\begin{itemize}[leftmargin=*]

\item In this study, we included 189 patches generated by 8 ASR tools to fix buggy code from 13 software projects. We believe this is a substantial number of patches generated by a substantial number of state-of-the-art ASR tools. Past empirical studies on ASR, e.g.,~\cite{qi2015analysis}, include five tools and 55 patches from 105 bugs. Still, we acknowledge that results may differ if more patches from more projects and more ASR tools are considered.
\item We have included 35 professional developers in our user study. The number of professional developers included in this study is larger or similar to those considered in many prior work, e.g.,~\cite{KevicWSSSF15, johnson2013don, rubin2016challenges}. Admittedly, it is possible that results differ for other groups of developers. To reduce this threat, we have selected the developers from two large IT companies and a large educational institution. We have also included a mix of junior and senior developers.
\end{itemize}


\noindent{\bf Threats to construct validity.} These threats relate to the suitability of our evaluation metrics. In this study, we use Krippendorff's alpha and average pairwise Cohen's kappa to evaluate the reliability of the patch labels from independent annotators. We also use the two to measure agreement between independent annotators' labels and those produced by author and automated annotations. These metrics are widely used in many research areas, e.g., information retrieval~\cite{castillo2006reference, meij2011combining, amigo2013general}, software engineering~\cite{chaparro2017detecting, abdalkareem2017developers}, etc. Thus, we believe there is little threat to construct validity.



\section{Related Work}\label{sec:related}
\hspace{-0.1cm} \textbf{Program repair.} We briefly discuss other repair techniques beyond the techniques  used in our study (e.g., GenProg~\cite{le2012systematic}, Kali~\cite{qi2015analysis}, Nopol~\cite{nopol:xuan}, and ACS~\cite{xiong2017precise}, etc), which have been described in Section~\ref{sec:background}. General program repair techniques can typically be divided into
two main branches: heuristic- and semantics-based repair. Heuristics-based
repair techniques
heuristically search for repairs commonly via genetic programming
algorithm. RSRepair~\cite{randomsearch} and AE~\cite{adaptivesearch} replace the
search strategy in GenProg by random and adaptive search strategies,
respectively. PAR~\cite{kim2013automatic} generates repairs
based on repair templates manually learned from human written patches. Prophet~\cite{long2016automatic} and HDRepair~\cite{le2016history} learn and mine repair
models from historical data for ranking patches, preferring those
that match frequent human fix patterns. Tan
\textit{et al.} propose anti-patterns to
prevent heuristics-based repair tools from generating
trivial repairs~\cite{antipatternrepair}.

Semantics-based repair techniques, such as SemFix~\cite{semfix},
DirectFix~\cite{mechtaev2015directfix}, and Angelix~\cite{mechtaev2016angelix},
synthesize repairs using symbolic execution and program synthesis. In a similar vein, S3~\cite{le2017s3} additionally proposes to employ various measures on the syntactic and semantics distances between candidates fixes and the original program to rank the search space. Other semantics-based techniques include
SPR~\cite{long2015staged}, which targets defects in if-conditions. Qlose~\cite{qlose} uses program execution
traces as an additional criteria to rank patches, and encode program repair
problem into a program synthesis tool namely SKETCH~\cite{sketchsynthesis}.
SearchRepair~\cite{Ke15ase} lies between heuristic- and semantic-based repair,
using semantic search as its underlying mutation approach to produce
higher-granularity, high-quality patches.  However, it does not yet scale as well as other approaches. Le et al. proposed to combine both search- and semantics-based repair into a single approach~\cite{le2016towards}.

\vspace{0.1cm}
\hspace{-0.35cm}\textbf{Empirical studies on patch correctness assessment.} To address patch correctness, two popular methods have been used, including author annotation and automated annotation via independent test suites generated by automatic test generation tools. Qi \textit{et al.}~\cite{qi2015analysis} empirically studied patches generated by GenProg~\cite{le2012systematic}, RSRepair~\cite{qi2014strength}, and AE~\cite{weimer2013leveraging}. They manually investigated the patches, wrote additional test cases, and reported the results on running the patches against additional test cases. Authors of PAR~\cite{kim2013automatic} performed a user study on the acceptability of patches generated by their tool. They employed 89 students and 164 developers to confirm that patches generated by PAR are more acceptable than GenProg. Monperrus \textit{et al.}~\cite{monperrus2014critical} discuss the main evaluation criteria of automatic software
repair including understandability, correctness and completeness. They suggest that repair techniques having their generated patches along with correctness labels kept private, such as PAR, are questionable. To avoid potential bias of manual human investigation, Smith \textit{et al.} use automatic test case generation tool KLEE~\cite{carterette2010effect} to generate independent test suites (ITS) that maximize coverage of ground-truth program to assess machine-generated patches~\cite{smith2015cure}. Using ITS, they evaluate the effectiveness of GenProg, RSRepair (aka. TrpAutoRepair), and AE on the IntroClass dataset~\cite{le2015manybugs} containing thousands of small programs. Our study is different from the mentioned studies in that we \emph{objectively} assess the reliability of author annotation and automated annotation.

\vspace{0.1cm}
\hspace{-0.35cm}\textbf{Empirical studies on biases and reliability.} A number of empirical studies have analyzed biases and reliability issues that affect how automated software engineering solutions are evaluated. Bird \textit{et al.} highlighted that only a fraction of bug fixes are labelled in version control systems and this causes a systematic bias in the evaluation of defect prediction tools~\cite{BirdBADBFD09}. Herzig \textit{et al.} manually examined 7,000 reports from issue tracking systems of open source projects and reported that 33.8\% of all bug reports to be misclassified~\cite{HerzigJZ13}. They showed that the misclassification introduces bias to defect prediction studies since a substantial number of files is wrongly marked as defective. The goal of our study is similar to the goals above mentioned studies -- we want to highlight and reduce bias in the evaluation of existing automated software engineering tools.

\vspace{-0.4cm}
\section{Conclusion and Future Work}\label{sec:conclusion}
In this paper, to assess reliability of existing patch correctness assessment methods, we conducted a user study with 35 professional developers to construct a gold set of correctness labels for 189 patches generated by different ASR techniques. By measuring inter-rater agreement (which was found to be substantial and on par with other high-quality benchmarks), we validated the quality of annotation labels in our gold set. We then compare our gold set with labels produced by authors (i.e., Liu et al.~\cite{liu2017identifying}, Martinez et al.~\cite{DBLP:journals/ese/MartinezDSXM17}, and Le et al.~\cite{le2017s3}) and independent test suites generated by \textsc{DiffTGen}~\cite{xin2017identifying} and \textsc{Randoop}~\cite{DBLP:conf/icse/PachecoLEB07}, and report their strengths and deficiencies. In particular, we find that a majority (88.8-89.0\%) of patch correctness labels generated by authors match those produced by independent annotators. On the other hand, only fewer than a fifth of incorrect patches can be labelled by independent test suites (ITSs) generated by {\sc DiffTGen} and \textsc{Randoop} as such. {\sc DiffTGen} and \textsc{Randoop} can however generate ITSs that can uncover multiple incorrect patches that are labeled as ``unknown'' or ``correct'' by authors. Based on our findings, we recommend that ASR authors release their patch correctness labels for public inspection. We also encourage more collaborative effort to distribute the expensive cost of ASR evaluation especially through user studies like ours. We also stressed that ITS alone should not be used to fully judge patch correctness labels; still, they can be used in conjunction with author annotation to help the latter produce labels that can more closely approximate independent annotators' labels.

In the future, we plan to expand our gold set by recruiting more professional developers and collecting more patches generated by additional ASR techniques through a large-scale collaborative effort among ASR researchers. We also plan to explore the possibility of organizing competitions with industrial data owners (e.g., with our two industrial partners whose developers have participated in this study) for further ASR research.  

\bibliographystyle{ACM-Reference-Format}
\bibliography{sample-bibliography}

\end{document}